\documentclass[a4paper,11pt]{article}
\usepackage{pos}
\usepackage{hyperref}
\title{Constraints on minimally and conformally coupled ultralight dark matter with the EPTA}

\author*[a]{Clemente Smarra}

\affiliation[a]{SISSA — International School for Advanced Studies,\\ Via Bonomea 265, 34136, Trieste, Italy and INFN, Sezione di Trieste}
\affiliation[a]{IFPU — Institute for Fundamental Physics of the Universe,\\ Via Beirut 2, 34014 Trieste, Italy}

\emailAdd{csmarra@sissa.it}

\abstract{Millisecond pulsars are extremely stable natural timekeepers. Pulsar Timing Array experiments, tracking subtle changes in the pulsars' rotation periods, can shed light on the presence of ultralight particles in our Galaxy. In this conference paper, we start by reviewing the most conservative scenario, in which ultralight particles interact only gravitationally. In this setting, we show that Pulsar Timing Arrays are able to constrain the presence of ultralight fields up to a few tenths of the observed dark matter abundance. Then, we consider conformally coupled ultralight candidates, demonstrating that the constraints on the universal scalar coupling of the field to Standard Model particles  improve on existing bounds by several orders of magnitude, in the relevant mass range analyzed by Pulsar Timing Arrays. The discussion presented here is based on \cite{Smarra:2023egv,Smarra:2024kvv}.}

\FullConference{
2nd General Meeting of COST Action COSMIC WISPers\\
3-6 September 2024\\
Istanbul, Turkey
}

\begin{document}
\maketitle

\section{Introduction}
The Cold Dark Matter (CDM) paradigm has been remarkably successful in explaining many aspects of the large-scale structure of the Universe. However, it encounters notable challenges below the kiloparsec scales. For instance, flat density profiles observed in the inner regions of galaxies are in tension with the steep power-law behavior predicted by pure CDM models (\textit{cusp-core problem})~\cite{Flores_1994, Moore_1994, Karukes_2015}. 
To address these issues, it may be argued that dark matter (DM) consists of ultralight scalar fields with masses $m \sim 10^{-22}~\text{eV}$ and negligible self-interactions~\cite{Hu:2000ke, Hui_2017}. The de Broglie wavelength of such fields in galaxies spans $\sim \text{kpc}$ scales, naturally suppressing small-scale power while preserving the CDM successes at larger scales.\\
In Ref.~\cite{Khmelnitsky_2014}, it was pointed out that the universal gravitational coupling of ultralight dark matter (ULDM) to ordinary matter influences the light travel time of radio signals emitted by pulsars.
Based on this principle, Pulsar Timing Array (PTA) experiments have established 95\% upper limits on the local energy density of ULDM, reaching $\rho\lesssim 0.15$ GeV/cm$^3$ in the mass range  $10^{-24.0}~\text{eV} \lesssim m \lesssim 10^{-23.7}~\text{eV}$~\cite{Smarra:2023egv, Porayko_2018}.\\
However, a natural possibility that respects the weak equivalence principle is that ULDM may be
universally (conformally) coupled to gravity, or (equivalently, in the Einstein frame) to the Standard Model.
In this context, ULDM may be regarded as a scalar-tensor theory of the Fierz-Jordan-Brans-Dicke~\cite{Fierz:1956zz,Jordan:1959eg,Brans_1961,Dicke62} or Damour-Esposito-Far\`{e}se type~\cite{Damour_1992,Damour_1993}, in the presence of a (light) mass potential term~\cite{Alsing:2011er}. As a result of the strong gravitational fields active inside neutron stars, the universal coupling to gravity produces a gravity-mediated interaction between neutron stars (and thus pulsars) and the scalar ULDM field~\cite{Nordtvedt68,Eardley1975ApJ,1989ApJ...346..366W,1977ApJ...214..826W,Damour_1992,Will:1993ns}.\\
This work is organized as follows. 
In Section~\ref{sec:min}, we introduce the Lagrangian of a minimally coupled ULDM field and provide a brief overview of key features relevant to our study. Then, we examine the impact of such a candidate on the times of arrival (TOAs) of pulsar radio beams, presenting bounds constraining the local ULDM abundance to values \textit{below} the measured local DM density in a specific mass range.  
In Section~\ref{sec:nonmin}, we analyze the case of a conformally coupled ULDM candidate. Consequently, we derive constraints on the coupling strength, achieving bounds that surpass those from Cassini tests of General Relativity~\cite{Bertotti_2003, Blas_2017} and observations of the pulsar in a triple stellar system~\cite{Ransom_2014, Archibald_2018, Voisin_2020} by several orders of magnitude within the mass range probed by pulsar timing arrays (PTAs).
Conclusions are summarized in Section~\ref{sec:conclusions}.

\section{Minimally coupled ULDM}\label{sec:min}
\begin{figure}[htbp]
\centering
\includegraphics[width=0.65\textwidth]{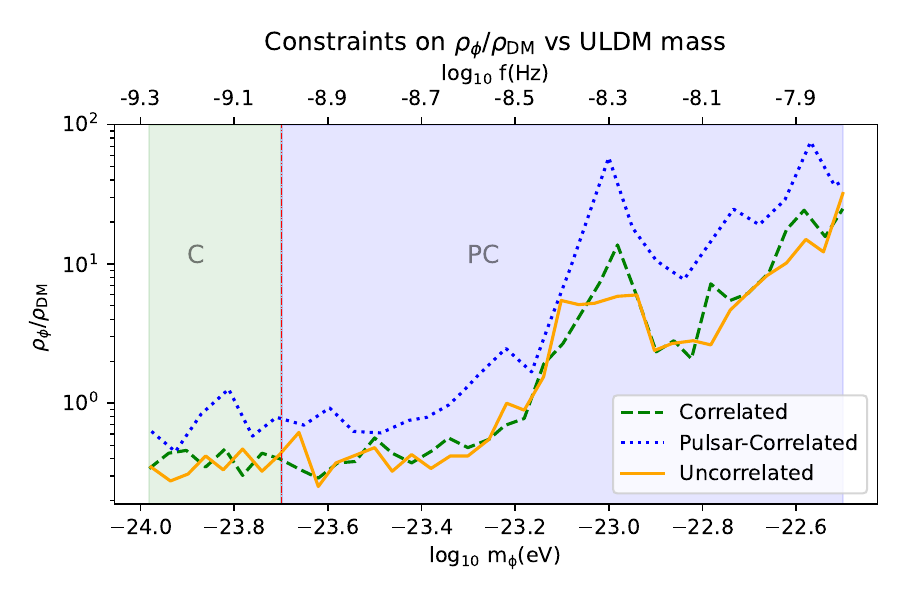}
\caption{Upper limits on the ratio of the scalar field density to the local abundance of DM, $\rho_\phi/\rho_\text{DM}$, at 95\% credibility versus the ULDM mass. We plot results for the \textit{uncorrelated} (U), \textit{pulsar-correlated} (PC) and \textit{correlated} (C) scenarios in solid, dotted and dashed lines, respectively. Moreover, we color-identify the regions of relevance of each of the three regimes. We truncate the plot at $\text{log}_\text{10}~m_\mathrm{\phi} (\text{eV})\sim -22.5$, as the upper bounds are not much informative anymore. As a result, the domain of validity of the uncorrelated limit is not shown here, and we refer to Ref.~\cite{Smarra:2023egv} for more details. The priors on the parameters of the analysis are summarized in Table I of Ref.~\cite{Smarra:2024kvv}.}\label{fig:rhomin}
\end{figure}

The action for an ultralight scalar field $\phi$ with negligible self-interactions and no couplings with the Standard Model is as simple as:
\begin{equation}
    S=\int d^4x\sqrt{-g}\left[\frac{1}{2}g^{\mu\nu}\partial_\mu\phi\partial_\nu\phi-\frac{1}{2}m_\phi^2\phi^2\right],
    \label{eq:action}
\end{equation}
where $m_\phi$ stands for the mass of the scalar field.
Given its high occupation number and non-relativistic nature, the ULDM scalar field can be treated as a classical wave~\cite{Khmelnitsky_2014}:
\begin{equation} 
\phi(\vec{x}, t)=\frac{\sqrt{2 \rho_\phi}}{m_\phi} \hat{\phi}(\vec{x}) \cos \left(m_\phi t+\gamma(\vec{x})\right), 
\label{eq:phi} 
\end{equation} 
where $\rho_\phi$ is the scalar field density, $\hat{\phi}(\vec{x})$ is a stochastic parameter, extracted from the Rayleigh distribution ($P(\hat{\phi}^2) = e^{-\hat{\phi}^2} $)~\cite{Castillo_2022} and encoding the interference pattern near $\vec{x}$ arising from the wave-like nature of ULDM, and $\gamma(\vec{x})$ is a spatially dependent phase. 
The ULDM scalar field is described by Eq.~\eqref{eq:phi} on timescales shorter than its \textit{coherence time} or, equivalently, on length scales smaller than its \textit{coherence length}
\begin{equation}
	\tau_c \sim  \frac{2}{mv^2} = 2\times 10^5~\text{yr} \left(\frac{10^{-22}~\text{eV}}{m}\right), \qquad l_c \sim \frac{1}{mv}  \sim 0.4~\text{kpc} \left(\frac{10^{-22}~\text{eV}}{m}\right),
	\label{eq:coh_time}
\end{equation}
where $v\sim 10^{-3}$ is the typical DM velocity in our Galaxy. 
The oscillating ULDM field leads to a periodic displacement $\delta t$ of the TOAs of radio pulses emitted by pulsars, which reads~\cite{Khmelnitsky_2014, Porayko_2018}:
\begin{equation}
    \delta t  = \frac{\rho_\phi}{2m_\phi^3 } [\hat{\phi}^2_\text{E}\sin{(2m_\phi t + \gamma_\text{E} )} - \hat{\phi}^2_\text{P}\sin{(2m_\phi t+ \gamma_\text{P} )} ], 
    \label{eq:st}
\end{equation} 
where $\gamma_\text{P} \equiv 2\gamma(\vec{x_\text{p}}) - 2 m_\phi d_p/c$ ($\gamma_\text{E} \equiv 2\gamma(\vec{x_\text{e}})$) is related to the phase of Eq.~\eqref{eq:phi} evaluated at the pulsar (Earth) location, with $d_p$ labeling the pulsar-Earth distance. Current uncertainties in pulsar distance measurements are of the order of $\mathcal{O}(0.1\div 1)~\text{kpc}$~\cite{Verbiest_2012}; therefore, these redefinitions produce effective pulsar-dependent random phases. 
Depending on the typical Earth-pulsar distance, we can distinguish three different scenarios: the \textit{correlated regime}, when both the typical inter-pulsar and pulsar-Earth separations and the typical Galacto-centric region probed by the most precise MW rotation curves measurements (approximately the inner $\sim 20$ kpc~\cite{Nesti_2013}) are smaller than the ULDM coherence length; the \textit{pulsar-correlated regime}, when the ULDM coherence length is larger than the inter-pulsar and pulsar-Earth separations, but does not extend to the typical Galacto-centric radius probed by rotation curves; the \textit{uncorrelated regime}, when the average inter-pulsar and pulsar-Earth separation is larger than the ULDM coherence length.
We present our results for the three regimes in Fig.~\ref{fig:rhomin}.

\section{Non minimally coupled ULDM}\label{sec:nonmin}
As a next step, we can explore scenarios in which the ultralight field is non minimally coupled to gravity.
Maintaining the same notation as Section~\ref{sec:min}, the action for a non minimally coupled field in the Einstein frame is expressed as~\footnote{Notice that, contrarily to Section~\ref{sec:min}, $\phi$ has a non canonical normalization, and appears in the action as an adimensional quantity multiplied by the Planck mass $M_{\text{P}}$. We stick to this convention, as it simplifies comparisons with gravitational phenomena.}:
\begin{align}
    S = M_{\text{P}}^2 \int \mathrm{d}^4x \sqrt{-g} \, &\bigg[ \frac{R}{2} - g^{\mu \nu} \partial_\mu \phi \partial_\nu \phi + m_\phi^2\phi^2 \bigg] \nonumber + S_m[ \psi_m, \tilde g_{\mu \nu}] \;.
    \label{eq:action}
\end{align}
The matter action $S_m$ incorporates a universal \textit{conformal} coupling of the scalar field to the matter content $\psi_m$ through the (Jordan) effective metric $g_{\mu \nu} = \mathcal{A}^2(\phi)$, with $\mathcal{A}^2(0) = 1$. 
Firstly, we specialize to the Fierz-Jordan-Brans-Dicke (FJBD) theory~\cite{Fierz:1956zz,Jordan:1959eg,Brans_1961,Dicke62}, which features a linear conformal coupling
\begin{equation}
    \mathcal{A}(\phi) = e^{\alpha \phi} \sim 1+\alpha \, \phi.
    \label{eq:fjbd}
\end{equation}
The Cassini mission provides a bound on the \textit{scalar coupling} $\alpha$ to the level of $\alpha^2\lesssim 10^{-5}$~\cite{Bertotti_2003}, while the triple  system PSR J0337+1715 provides a more stringent limit of $\alpha^2\lesssim 4\times10^{-6}$~\cite{Voisin_2020}.\\
Secondly, we focus on the Damour-Esposito-Far\`{e}se (DEF) theory~\cite{Damour_1992,Damour_1993}, with a quadratic universal coupling:
\begin{equation}
    \mathcal{A}(\phi) = e^{\beta\phi^2/2}.
    \label{eq:def}
\end{equation}
The non-observation of deviations from the General Relativity (GR) predictions in binary pulsar data imposes a constraint on $\beta$, requiring $\beta \gtrsim -4.3$ (depending on the specific equation of state (EoS) for the neutron star model), to prevent non-perturbative spontaneous scalarization phenomena~\cite{Damour_1993, Shao:2017gwu}. 
In these models, the analogous of Eq.~\eqref{eq:phi} reads~\footnote{The factor of 2 of difference in the numerator arises from the non canonical normalization of the field $\phi$ in Eq.~\eqref{eq:action}.}:
\begin{equation} 
\phi(\vec{x}, t)=\frac{\sqrt{\rho_\phi}}{m_\phi M_P} \hat{\phi}(\vec{x}) \cos \left(m_\phi t+\gamma(\vec{x})\right). 
\label{eq:scal} 
\end{equation} 
In the Jordan frame, an oscillating scalar field, as presented in Eq.~\eqref{eq:scal}, leads to a temporal variation of Newton's constant, which affects the gravitational mass and radius of the neutron star~\cite{Damour_1992}. This dependence is captured by the sensitivities, defined as:
\begin{equation} \label{eq:angularmomsensitivity}
    s_I = - \frac{1}{2 \alpha(\phi)} \frac{\mathrm{d} \ln I}{\mathrm{d} \phi} \bigg |_{N,J} = \frac{1}{2 \alpha(\phi)} \frac{\mathrm{d} \ln \Omega_\mathrm{obs}}{\mathrm{d} \phi} \bigg |_{N,J},
\end{equation}
evaluated at constant pulsar's baryon number $N$ and Einstein-frame angular momentum $J$, where  $\alpha(\phi) = \mathrm{d} \log \mathcal{A} / \mathrm{d} \phi$. We use the code from Ref.~\cite{Kuntz_2024} to compute $s_I$.
With this relation at hand, a variation in the scalar field value directly reflects to a change in the pulsar's spin frequency, which in turn affects the TOAs.
\subsection{FJBD gravity theory}
\begin{figure}[htbp]
\centering
\includegraphics[width=0.6\textwidth]{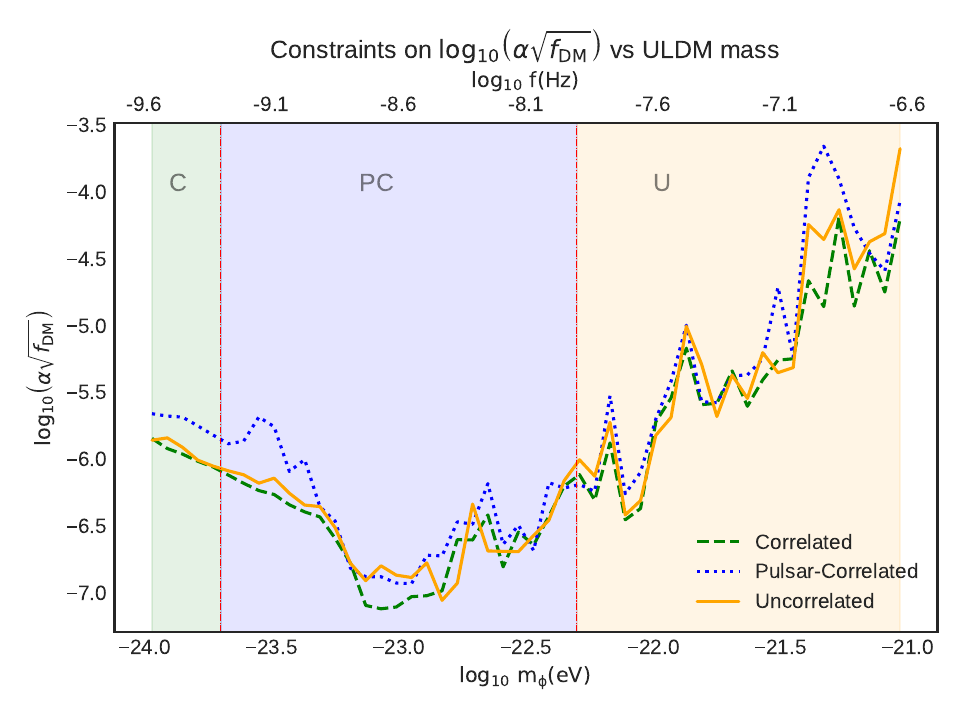}
\caption{Upper limits on $\log_{10}\left(\alpha\sqrt{f_\text{DM}}\right)$ at 95\% credibility versus the ULDM mass. We plot results for the \textit{uncorrelated} (U), \textit{pulsar-correlated} (PC) and \textit{correlated} (C) scenarios in solid, dotted and dashed lines, respectively. Moreover, we color-identify the regions of relevance of each of the three regimes. To produce the results, we used the AP4 EoS~\cite{Akmal_1998}, and the priors on the parameters of the analysis are summarized in Table I of Ref.~\cite{Smarra:2024kvv}.}\label{fig:comp_tot}
\end{figure}

In this scenario, inspecting Eq.~\eqref{eq:fjbd} and using the definition of $\alpha(\phi)$, we find that $\alpha(\phi) = \alpha$, where $\alpha$ is constant. Furthermore, the numerical analysis conducted in Ref.~\cite{Kuntz_2024} demonstrates that the angular momentum sensitivity $s_I$ exhibits only a very weak dependence on the scalar coupling $\alpha$. Given this, we neglect this dependence in our analysis.\\
In analogy to Eq.~\eqref{eq:st}, the timing residuals in this context are given by:
\begin{equation}
    \delta t = \left.2\alpha\frac{\sqrt{\rho_\phi}}{M_\text{P} m_\phi^2} s_I \hat{\phi}_\text{P}(\vec{x}) \sin(m_\phi t + \gamma(\vec{x}))\right|_{t_{\text {start }}} ^{t_{\text {end }}}.
    \label{eq:dt}
\end{equation}
We refer the reader to Ref.~\cite{Smarra:2024kvv} for details on the derivation of this result. Defining $f_\text{DM} \equiv \rho_\phi/\rho_\text{DM}$ as the fraction of the total abundance of DM, assumed to be fiducially $\rho_\text{DM} = 0.4~\text{GeV/cm}^3$, our ULDM candidate accounts for, we can present results in terms of $\alpha\sqrt{f_\text{DM}}$.
This is indeed the quantity constrained by Eq.~\eqref{eq:dt}, enabling a straightforward determination of the bound on $\alpha$ once a specific value for the scalar field density $\rho_\phi$ is assumed. 
Fig.~\ref{fig:comp_tot} shows the upper limits for the \textit{correlated, pulsar correlated} and \textit{uncorrelated} scenarios. 

\subsection{DEF gravity theory}
\begin{figure}
	\centering
    \includegraphics[width = 0.495\textwidth]{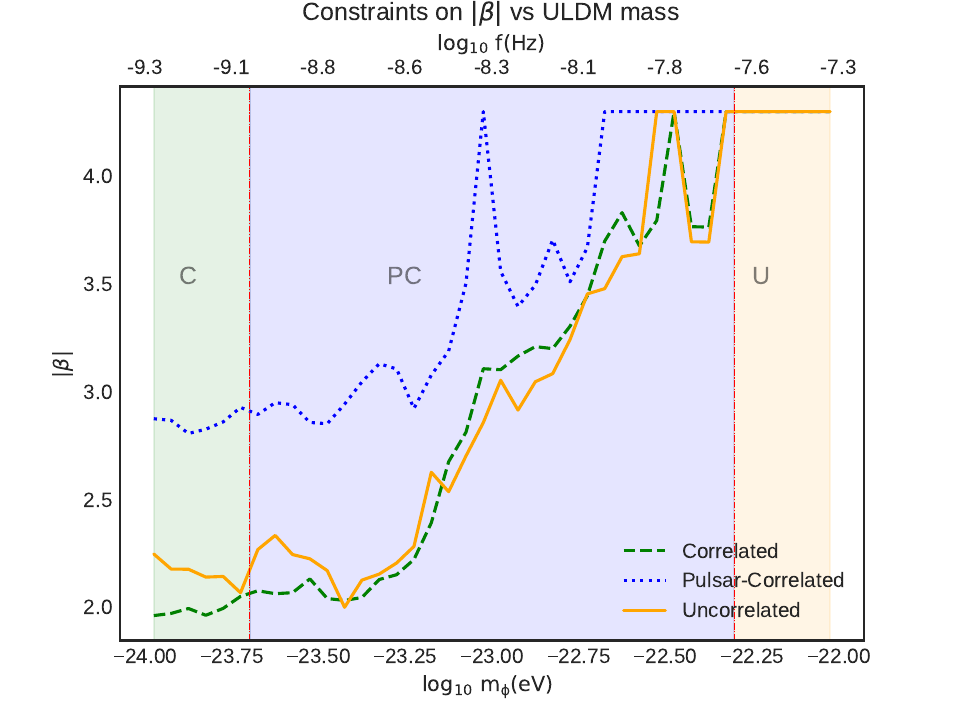}
    \includegraphics[width = 0.475\textwidth]{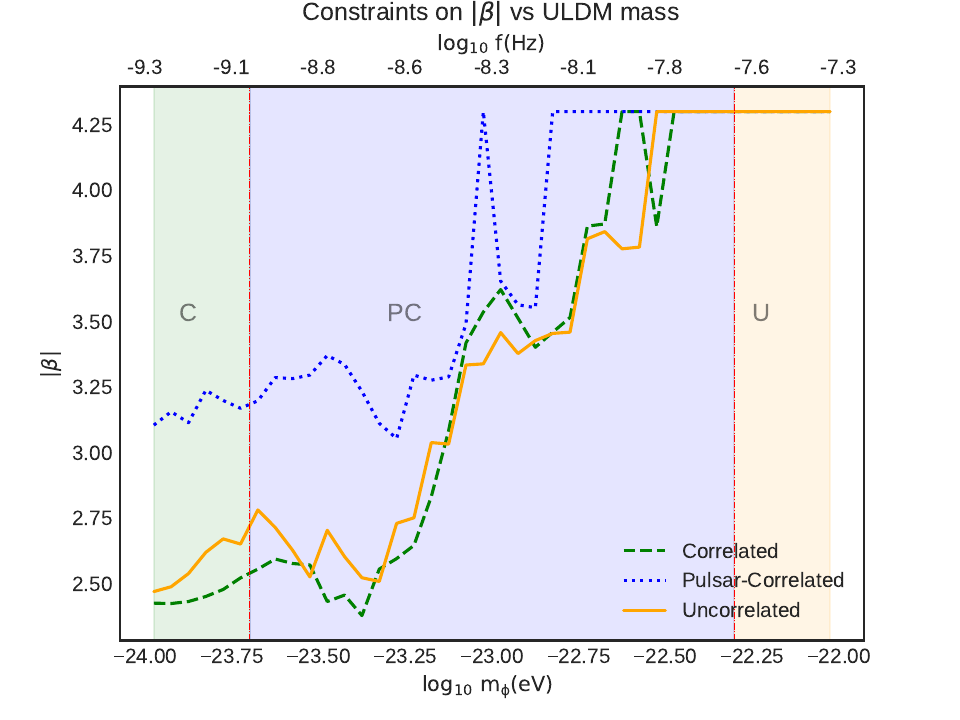}\\
    \caption{Upper limits on $\lvert\beta\rvert$  ($\beta < 0$) at 95\% credibility versus the ULDM mass. We show bounds assuming  $\rho = \rho_\text{DM}$ and  $\rho =0.5~\rho_\text{DM}$ in the left and right panels, respectively. We plot limits for the \textit{uncorrelated} (U), \textit{pulsar-correlated} (PC) and \textit{correlated} (C)  scenarios in solid, dotted and dashed lines, respectively. Moreover, we color-identify the regions of relevance of each of the three regimes. To produce the results, we used the MPA1 EoS~\cite{Muther_1987}. In cases where the data are not constraining, the upper limits correspond to the maximum value allowed by our prior. The priors for the parameters relevant to this analysis are defined according to the scheme outlined in Table I of Ref.~\cite{Smarra:2024kvv}.
    }
    \label{fig:beta_neg}
\end{figure}

In this case, by looking at Eq.~\eqref{eq:def}, we have $\alpha(\phi) = \beta\phi$. In contrast to the FJBD scenario, here the angular momentum sensitivity depends non trivially on the coupling $\beta$. Therefore, we write $s_I = s_I(\beta)$ to keep this in mind.
The expression for the timing residuals is given by
\begin{equation}
    \delta t = \left.\frac{\rho_\phi}{2m_\phi^3 M_P^2} \beta s_I (\beta) \hat{\phi}_\text{P}^2(\vec{x}) \sin(2m_\phi t + \gamma(\vec{x}))\right|_{t_{\text {start }}} ^{t_{\text {end }}}.
    \label{eq:dtbeta}
\end{equation}
The ULDM signal~\eqref{eq:dtbeta} depends separately on $\rho_\phi$ and $\beta$, which must therefore be treated as two independent parameters in the analysis.\\
To address this, we select two benchmark values for the scalar field density, namely $\rho_\phi = \rho_\text{DM}$ and $\rho_\phi = 0.5~ \rho_\text{DM}$, and we present the resulting bounds on $\beta$ in Fig.~\ref{fig:beta_neg}.\\
Here, we restrict to negative values of $\beta$, and we choose $\beta = -4.3$ as the lower bound for the prior, as more negative values would induce $\mathcal{O}(1)$ deviations from GR and are excluded by binary pulsars~\cite{Shao:2017gwu}.  We refer the reader to Ref.\cite{Smarra:2024kvv} for an extended discussion of this model, which includes also positive values of $\beta$. 

\section{Conclusions}\label{sec:conclusions} 
In this work, we reviewed the constraints that the EPTA sets on scalar ULDM density and couplings to gravity~\cite{Smarra:2023egv, Smarra:2024kvv}. In the minimally coupled case, scalar ULDM is allowed to contribute at most up to a few tenths of the observed DM abundance in the mass range $m_\phi \in [10^{-24}~\text{eV}, 10^{-23.7}~\text{eV}]$, in order to escape detection. In the non minimally coupled scenario, we specialized to the FJBD and the DEF gravity theories, and we considered the effects of a conformally coupled ULDM candidate on the observed TOAs. In the FJBD, our bounds improve by several orders of magnitude previous constraints, set by the Cassini spacecraft mission~\cite{Bertotti_2003} and the PSR J0337+1715 system~\cite{Voisin_2020}, across the entire mass range the EPTA is sensitive to. In a specific mass range, our results yield improved bounds compared to the previous literature in the DEF case as well~\cite{Shao:2017gwu, Mendes:2014ufa, Anderson:2017phb}. We also refer the interested reader to Refs.~\cite{Kaplan_2022, wu:2024} for an extended discussion on how to constrain ULDM couplings to the Standard Model and to Ref.~\cite{EPTA:2024gxu}  for an interesting approach on how to constrain the ULDM coupling to photons using  polarization data.

\section*{Acknowledgements}
I wish to thank the organizers of the 2nd General Meeting of the COST Action CA21106 for setting up a truly wonderful and exciting meeting. This article is based on the work from COST Action COSMIC WISPers CA21106, supported by COST (European Cooperation in
Science and Technology).


\begin{thebibliography}{99}
\bibitem{Smarra:2023egv}
{\scshape European Pulsar Timing Array} collaboration, C.~Smarra et~al.,
  \emph{{Second Data Release from the European Pulsar Timing Array: Challenging
  the Ultralight Dark Matter Paradigm}},
  \href{https://doi.org/10.1103/PhysRevLett.131.171001}{\emph{Phys. Rev. Lett.}
  {\bfseries 131} (2023) 171001},
  [\href{https://arxiv.org/abs/2306.16228}{{\ttfamily 2306.16228}}].

\bibitem{Smarra:2024kvv}
C.~Smarra et~al., \emph{{Constraints on conformal ultralight dark matter
  couplings from the European Pulsar Timing Array}},
  \href{https://doi.org/10.1103/PhysRevD.110.043033}{\emph{Phys. Rev. D}
  {\bfseries 110} (2024) 043033},
  [\href{https://arxiv.org/abs/2405.01633}{{\ttfamily 2405.01633}}].

\bibitem{Flores_1994}
R.~A. {Flores} and J.~R. {Primack}, \emph{{Observational and Theoretical
  Constraints on Singular Dark Matter Halos}},
  \href{https://doi.org/10.1086/187350}{\emph{Astrophysical Journal, Letters} (May, 1994) L1
  {\bfseries 427} }
\bibitem{Moore_1994}
B.~Moore, \emph{Evidence against dissipation-less dark matter from observations
  of galaxy haloes}, \href{https://doi.org/10.1038/370629a0}{\emph{Nature}
  {\bfseries 370} (Aug, 1994) 629--631}.

\bibitem{Karukes_2015}
{Karukes, E. V.}, {Salucci, P.} and {Gentile, G.}, \emph{The dark matter
  distribution in the spiral ngc 3198 out to 0.22 rvir},
  \href{https://doi.org/10.1051/0004-6361/201425339}{\emph{A\&A} {\bfseries
  578} (2015) A13}.

\bibitem{Hu:2000ke}
W.~Hu, R.~Barkana and A.~Gruzinov, \emph{{Cold and fuzzy dark matter}},
  \href{https://doi.org/10.1103/PhysRevLett.85.1158}{\emph{Phys. Rev. Lett.}
  {\bfseries 85} (2000) 1158--1161},
  [\href{https://arxiv.org/abs/astro-ph/0003365}{{\ttfamily
  astro-ph/0003365}}].

\bibitem{Hui_2017}
L.~Hui, J.~P. Ostriker, S.~Tremaine and E.~Witten, \emph{Ultralight scalars as
  cosmological dark matter},
  \href{https://doi.org/10.1103/physrevd.95.043541}{\emph{Physical Review D}
  {\bfseries 95} (feb, 2017) }.


\bibitem{Khmelnitsky_2014}
A.~Khmelnitsky and V.~Rubakov, \emph{Pulsar timing signal from ultralight
  scalar dark matter},
  \href{https://doi.org/10.1088/1475-7516/2014/02/019}{\emph{Journal of
  Cosmology and Astroparticle Physics} {\bfseries 2014} (feb, 2014) 019--019}.


\bibitem{Porayko_2018}
N.~K. Porayko, X.~Zhu, Y.~Levin, L.~Hui, G.~Hobbs, A.~Grudskaya et~al.,
  \emph{Parkes pulsar timing array constraints on ultralight scalar-field dark
  matter}, \href{https://doi.org/10.1103/physrevd.98.102002}{\emph{Physical
  Review D} {\bfseries 98} (Nov., 2018) }.

  \bibitem{Fierz:1956zz}
M.~Fierz, \emph{{On the physical interpretation of P.Jordan's extended theory
  of gravitation}}, {\emph{Helv.\ Phys.\ Acta} {\bfseries 29} (1956) 128--134}.

\bibitem{Jordan:1959eg}
P.~Jordan, \emph{{The present state of Dirac's cosmological hypothesis}},
  \href{https://doi.org/10.1007/BF01375155}{\emph{Z.\ Phys.} {\bfseries 157}
  (1959) 112--121}.

\bibitem{Brans_1961}
C.~Brans and R.~H. Dicke, \emph{Mach's principle and a relativistic theory of
  gravitation}, \href{https://doi.org/10.1103/PhysRev.124.925}{\emph{Phys.
  Rev.} {\bfseries 124} (Nov, 1961) 925--935}.

\bibitem{Dicke62}
R.~H. Dicke, \emph{Mach's principle and invariance under transformation of
  units}, \href{https://doi.org/10.1103/PhysRev.125.2163}{\emph{Phys. Rev.}
  {\bfseries 125} (Mar, 1962) 2163--2167}.

\bibitem{Damour_1992}
T.~Damour and G.~Esposito-Farese, \emph{{Tensor multiscalar theories of
  gravitation}}, \href{https://doi.org/10.1088/0264-9381/9/9/015}{\emph{Class.
  Quant. Grav.} {\bfseries 9} (1992) 2093--2176}.

\bibitem{Damour_1993}
T.~Damour and G.~Esposito-Far\`ese, \emph{Nonperturbative strong-field effects
  in tensor-scalar theories of gravitation},
  \href{https://doi.org/10.1103/PhysRevLett.70.2220}{\emph{Phys. Rev. Lett.}
  {\bfseries 70} (Apr, 1993) 2220--2223}.

\bibitem{Alsing:2011er}
J.~Alsing, E.~Berti, C.~M. Will and H.~Zaglauer, \emph{{Gravitational radiation
  from compact binary systems in the massive Brans-Dicke theory of gravity}},
  \href{https://doi.org/10.1103/PhysRevD.85.064041}{\emph{Phys. Rev. D}
  {\bfseries 85} (2012) 064041},
  [\href{https://arxiv.org/abs/1112.4903}{{\ttfamily 1112.4903}}].

\bibitem{Nordtvedt68}
K.~Nordtvedt, \emph{Equivalence principle for massive bodies. ii. theory},
  \href{https://doi.org/10.1103/PhysRev.169.1017}{\emph{Phys. Rev.} {\bfseries
  169} (May, 1968) 1017--1025}.

\bibitem{Eardley1975ApJ}
D.~M. {Eardley}, \emph{{Observable effects of a scalar gravitational field in a
  binary pulsar}}, \href{https://doi.org/10.1086/181744}{\emph{Astrophysical
  Journal} {\bfseries 196} (Mar., 1975) L59--L62}.

\bibitem{1989ApJ...346..366W}
C.~M. {Will} and H.~W. {Zaglauer}, \emph{{Gravitational Radiation, Close Binary
  Systems, and the Brans-Dicke Theory of Gravity}},
  \href{https://doi.org/10.1086/168016}{\emph{Astrophysical Journal} {\bfseries 346} (Nov.,
  1989) 366}.

\bibitem{1977ApJ...214..826W}
C.~M. {Will}, \emph{{Gravitational radiation from binary systems in alternative
  metric theories of gravity: dipole radiation and the binary pulsar.}},
  \href{https://doi.org/10.1086/155313}{\emph{Astrophysical Journal} {\bfseries 214} (June,
  1977) 826--839}.

\bibitem{Will:1993ns}
C.~M. Will, \emph{{Theory and experiment in gravitational physics}}.
\newblock 1993.


\bibitem{Bertotti_2003}
B.~Bertotti, L.~Iess and P.~Tortora, \emph{A test of general relativity using
  radio links with the cassini spacecraft},
  \href{https://doi.org/10.1038/nature01997}{\emph{Nature} {\bfseries 425}
  (Sep, 2003) 374--376}.

  \bibitem{Blas_2017}
D.~Blas, D.~L. Nacir and S.~Sibiryakov, \emph{Ultralight dark matter resonates
  with binary pulsars},
  \href{https://doi.org/10.1103/physrevlett.118.261102}{\emph{Physical Review
  Letters} {\bfseries 118} (jun, 2017) }.


\bibitem{Ransom_2014}
S.~M. Ransom, I.~H. Stairs, A.~M. Archibald, J.~W.~T. Hessels, D.~L. Kaplan,
  M.~H. van Kerkwijk et~al., \emph{A millisecond pulsar in a stellar triple
  system}, \href{https://doi.org/10.1038/nature12917}{\emph{Nature} {\bfseries
  505} (Jan, 2014) 520--524}.

\bibitem{Archibald_2018}
A.~M. Archibald, N.~V. Gusinskaia, J.~W.~T. Hessels, A.~T. Deller, D.~L.
  Kaplan, D.~R. Lorimer et~al., \emph{Universality of free fall from the
  orbital motion of a pulsar in a stellar triple system},
  \href{https://doi.org/10.1038/s41586-018-0265-1}{\emph{Nature} {\bfseries
  559} (Jul, 2018) 73--76}.

\bibitem{Voisin_2020}
{Voisin, G.}, {Cognard, I.}, {Freire, P. C. C.}, {Wex, N.}, {Guillemot, L.},
  {Desvignes, G.} et~al., \emph{An improved test of the strong equivalence
  principle with the pulsar in a triple star system},
  \href{https://doi.org/10.1051/0004-6361/202038104}{\emph{A\& A} {\bfseries
  638} (2020) A24}.

\bibitem{Castillo_2022}
A.~Castillo, J.~Martin-Camalich, J.~Terol-Calvo, D.~Blas, A.~Caputo, R.~T.~G.
  Santos et~al., \emph{{Searching for dark-matter waves with PPTA and QUIJOTE
  pulsar polarimetry}},
  \href{https://doi.org/10.1088/1475-7516/2022/06/014}{\emph{JCAP} {\bfseries
  06} (2022) 014}, [\href{https://arxiv.org/abs/2201.03422}{{\ttfamily
  2201.03422}}].

\bibitem{Verbiest_2012}
J.~P.~W. Verbiest, J.~M. Weisberg, A.~A. Chael, K.~J. Lee and D.~R. Lorimer,
  \emph{On pulsar distance measurements and their uncertainties},
  \href{https://doi.org/10.1088/0004-637X/755/1/39}{\emph{The Astrophysical
  Journal} {\bfseries 755} (jul, 2012) 39}.

\bibitem{Nesti_2013}
F.~Nesti and P.~Salucci, \emph{The dark matter halo of the milky way, ad 2013},
  \href{https://doi.org/10.1088/1475-7516/2013/07/016}{\emph{Journal of
  Cosmology and Astroparticle Physics} {\bfseries 2013} (jul, 2013) 016}.


\bibitem{Kuntz_2024}
A.~Kuntz and E.~Barausse, \emph{Angular momentum sensitivities in scalar-tensor
  theories}, \href{https://doi.org/10.1103/PhysRevD.109.124001}{\emph{Phys.
  Rev. D} {\bfseries 109} (Jun, 2024) 124001}.
  
\bibitem{Akmal_1998}
A.~Akmal, V.~R. Pandharipande and D.~G. Ravenhall, \emph{Equation of state of
  nucleon matter and neutron star structure},
  \href{https://doi.org/10.1103/PhysRevC.58.1804}{\emph{Phys. Rev. C}
  {\bfseries 58} (Sep, 1998) 1804--1828}.

  \bibitem{Muther_1987}
H.~Müther, M.~Prakash and T.~Ainsworth, \emph{The nuclear symmetry energy in
  relativistic brueckner-hartree-fock calculations},
  \href{https://doi.org/https://doi.org/10.1016/0370-2693(87)91611-X}{\emph{Physics
  Letters B} {\bfseries 199} (1987) 469--474}.

  \bibitem{Shao:2017gwu}
L.~Shao, N.~Sennett, A.~Buonanno, M.~Kramer and N.~Wex, \emph{{Constraining
  nonperturbative strong-field effects in scalar-tensor gravity by combining
  pulsar timing and laser-interferometer gravitational-wave detectors}},
  \href{https://doi.org/10.1103/PhysRevX.7.041025}{\emph{Phys. Rev. X}
  {\bfseries 7} (2017) 041025},
  [\href{https://arxiv.org/abs/1704.07561}{{\ttfamily 1704.07561}}].

\bibitem{Mendes:2014ufa}
R.~F.~P. Mendes, \emph{{Possibility of setting a new constraint to
  scalar-tensor theories}},
  \href{https://doi.org/10.1103/PhysRevD.91.064024}{\emph{Phys. Rev. D}
  {\bfseries 91} (2015) 064024},
  [\href{https://arxiv.org/abs/1412.6789}{{\ttfamily 1412.6789}}].

\bibitem{Anderson:2017phb}
D.~Anderson and N.~Yunes, \emph{{Solar System constraints on massless
  scalar-tensor gravity with positive coupling constant upon cosmological
  evolution of the scalar field}},
  \href{https://doi.org/10.1103/PhysRevD.96.064037}{\emph{Phys. Rev. D}
  {\bfseries 96} (2017) 064037},
  [\href{https://arxiv.org/abs/1705.06351}{{\ttfamily 1705.06351}}].

  
\bibitem{Kaplan_2022}
D.~E. Kaplan, A.~Mitridate and T.~Trickle, \emph{Constraining fundamental
  constant variations from ultralight dark matter with pulsar timing arrays},
  \href{https://doi.org/10.1103/PhysRevD.106.035032}{\emph{Phys. Rev. D}
  {\bfseries 106} (Aug, 2022) 035032}.

\bibitem{wu:2024}
Y.-M. Wu and Q.-G. Huang, \emph{Constraining ultralight scalar dark matter
  couplings with the european pulsar timing array second data release},  2024.

  
\bibitem{EPTA:2024gxu}
{\scshape EPTA} collaboration, N.~K. Porayko et~al., \emph{{Searches for
  signatures of ultra-light axion dark matter in polarimetry data of the
  European Pulsar Timing Array}},
  \href{https://arxiv.org/abs/2412.02232}{{\ttfamily 2412.02232}}.

\end{thebibliography}
\end{document}